\documentclass[a4paper,8pt]{article}

\usepackage{fancyhdr}

\usepackage{multicol}
\usepackage{etoolbox}
\usepackage{cancel}
\usepackage[table]{xcolor}

\usepackage[font=small,labelfont=bf]{caption}
\DeclareCaptionFont{scriptsize}{\scriptsize}
\DeclareCaptionFont{tiny}{\tiny}
\usepackage{float}
\usepackage{amsmath, amssymb, setspace,cite,verbatim}
\usepackage{graphicx}
\usepackage{hyperref}
\usepackage{color}
\usepackage{wrapfig}

\usepackage{multicol}
\usepackage{etoolbox}
\usepackage{relsize}
\usepackage{multirow}

\usepackage{latexsym}
\usepackage{a4wide}
\usepackage{graphicx, subfigure}
\usepackage{cite}
\usepackage{tabularx}
\usepackage{array}
\usepackage{graphics}     
\usepackage{amsmath}
\usepackage{amssymb}
\usepackage{longtable} 
\usepackage{verbatim} 
\usepackage[table]{xcolor}
\usepackage{booktabs}
\usepackage{hyperref} 
\usepackage[utf8]{inputenc}
\usepackage{soul}

\usepackage{amsmath}
\usepackage{amssymb}

\newcommand{\fms}[1]{{#1}\!\!\!\slash}

\newcommand{\nb}{\overline{n}}

\newcommand{\nbs}{\fms{\overline{n}}}

\newcommand{\ie}{i\epsilon}

\newcommand{\braB}{\,\langle B \!\mid\,}
\newcommand{\ketB}{\,\mid\! B \rangle\,}

\newcommand{\nn}{\nonumber}

\usepackage{slashed}

\usepackage{pifont}

\definecolor{light-gray}{gray}{0.90}

\begin{document}

\hfill {\tt  MITP/20-014, Desy 20-084}  

\def\thefootnote{\fnsymbol{footnote}}
 
\begin{center}

\vspace{3.cm}

{\Large\bf {Resolved $1/m_b$ contributions to $\bar B \to X_{s,d} \ell^+ \ell^-$ and $\bar B \to X_s \gamma$} }

\setlength{\textwidth}{11cm}
                    
\vspace{2.cm}
{\large\bf  
Michael Benzke$^{\,a}$\footnote{Email: michael.benzke@desy.de},    
Tobias Hurth$^{b}$\footnote{Email: tobias.hurth@cern.ch},
}  
 
\vspace{1.cm}
{\em $^a$II. Institute for Theoretical Physics, University Hamburg\\ 
Luruper Chaussee 149, D-26761 Hamburg, Germany}\\[0.2cm]
{\em $^b$PRISMA+ Cluster of Excellence and  Institute for Physics (THEP)\\
Johannes Gutenberg University, D-55099 Mainz, Germany}\\[0.2cm] 

\end{center}

\renewcommand{\thefootnote}{\arabic{footnote}}
\setcounter{footnote}{0}

\vspace{1.cm}
\thispagestyle{empty}
\centerline{\bf ABSTRACT}
\vspace{0.5cm}
In view of the importance of the nonperturbative resolved contributions 
for the overall uncertainties of the two inclusive penguin decays $\bar B \to X_s \gamma$ and $\bar B \to X_{s,d} \ell^+\ell^-$ we 
reanalyse these contributions using new estimates of moments of the subleading shape functions and of other input parameters. 
Within a systematic approach we find {a} significant reduction of the nonperturbative uncertainties in the inclusive 
decay $\bar B \to X_{s,d} \ell^+\ell^-$, but  a much less pronounced reduction in the inclusive decay $\bar B \to X_s \gamma$ compared 
to a recent analysis on the resolved contributions to the inclusive decay $\bar B \to X_s \gamma$. We identify the reasons for this discrepancy.

\clearpage

\section{Introduction and new inputs}
The so-called resolved contributions to rare $B$-decays are non-local power corrections and can be systematically calculated using soft-collinear effective theory (SCET). In case of the inclusive $\bar B \to X_{s} \gamma$ decays all resolved contibutions to $O(1/m_b)$ have been analysed some time ago~\cite{Benzke:2010js,Benzke:2010tq,Lee:2006wn}. Also the analogous contributions to the inclusive $\bar B \to X_{s ,d} \ell^+ \ell^-$ decays have been calculated to $O(1/m_b)$~\cite{Hurth:2017xzf,Benzke:2017woq}. In both cases these analyses lead to an additional uncertainty of $4-5\%$ which represents the largest uncertainty in the prediction of the decay rate of $\bar B \to X_{s} \gamma$~\cite{Misiak:2015xwa} and of the low-$q^2$ observables of $\bar B \to X_{s ,d} \ell^+ \ell^-$~\cite{Huber:2015sra,Huber:2019iqf}. The resolved contributions contain subprocesses in which the photon couples to light partons instead of connecting directly to the effective weak-interaction vertex.
In both cases there are four contributions at $O(1/m_b)$, namely from the interference terms ${\cal O}_{7\gamma} - {\cal O}_{8g}$,\, ${\cal O}_{8g} - {\cal O}_{8g}$, and ${\cal O}^{c}_{1} - {\cal O}_{7\gamma}$, but also from ${\cal O}^{u}_1 - {\cal O}_{7\gamma}$. The latter is CKM suppressed in the $b \to s$ case, but was shown to vanish~\cite{Benzke:2010js}. It turns out that the ${\cal O}^{c}_{1} - {\cal O}_{7\gamma}$ piece has the largest impact. 
The resolved contributions are given by convolution integrals of a so-called jet-function, characterising the hadronic final state $X_{s(d)}$ at the intermediate hard-collinear scale $\sqrt{m_b \Lambda_{\rm QCD}}$, and of a soft (shape) function at scale $\Lambda_{\rm QCD}$ which is defined by an explicit non-local heavy-quark effective theory (HQET) matrix element. The hard contribution at the scale $m_b$ is factorised into Wilson coefficients.
The resolved contributions in the $\bar B \to X_{s ,d} \ell^+ \ell^-$ were calculated in the presence of a cut in the hadronic mass $M_X$; such a cut might be necessary also at the Belle-II experiment in order to suppress huge background from double semi-leptonic decays. However, it was explicitly shown~\cite{Hurth:2017xzf, Benzke:2017woq} that the resolved contributions stay nonlocal when the hadronic cut is released and, thus, represent an irreducible uncertainty. The support properties of the shape function imply that the resolved contributions (besides the ${\cal O}_{8g} - {\cal O}_{8g}$ one) are almost cut-independent.

The resolved contributions can be estimated in a conservative way by considering the explicit form of the HQET matrix element which represents the shape function. One can derive general properties of that matrix element and then use functions fulfilling all these properties in the convolution with the perturbatively calculated jet function to estimate the impact of the resolved contributions. In a recent paper~\cite{Gunawardana:2019gep}, {new estimates of}  the moments of the subleading shape function in the interference term ${\cal O}^{c}_{1} - {\cal O}_{7\gamma}$ -- based on the results in Refs.~\cite{Gunawardana:2017zix,Gambino:2016jkc} -- were derived and used to significantly reduce the uncertainty due to this resolved contribution in the decay $\bar B \to X_s \gamma$. In the present paper we revise our analysis of this resolved contribution to $\bar B \to X_{s ,d} \ell^+ \ell^-$ in view of these new estimates of the moments. 
{In our revised analysis we 
analyse all parametric uncertainties of input parameters and also  the scale dependence of our results} in order to get a reasonable estimate of this contribution in both inclusive decay modes. In the original analysis of the $\bar B \to X_s \gamma$  case~\cite{Benzke:2010js,Benzke:2010tq} often just central values of input parameters were used and scale dependences were not considered.

In the present analysis we follow the original choice in Ref.~\cite{Benzke:2010js} for the bottom quark and use the low-scale subtracted heavy quark mass defined in the shape function scheme~\cite{Bosch:2004th}. As in the new analysis in Ref.~\cite{Gunawardana:2019gep} we choose the latest HFLAV determination of that mass~\cite{Amhis:2016xyh}, namely $m_b=(4.58 \pm 0.03)\,\text{GeV}$. In comparison the original analysis in Ref.~\cite{Benzke:2010js} used the central value of $m_b=4.65\, \text{GeV}$ {and neglected any uncertainties.} 

{The charm mass dependence originates from the charm penguin diagram with a soft gluon emission in the ${\cal O}^{c}_{1} - {\cal O}_{7\gamma}$ interference term which is naturally calculated at the hard-collinear scale.
Thus, it is appropriate to consider the running charm mass at the hard-collinear scale ${m}_c^{\rm MS}(\mu_{\rm hc} )$. In order to make the ambiguity of the charm mass manifest, we change the hard-collinear scale $\mu_{\rm hc} \sim \sqrt{m_b\,\Lambda_{\rm QCD}}$  from $1.3\, {\rm GeV}$  to   $1.7\, {\rm GeV}$.} With the present PDG value of the charm mass being ${m}_c^{\rm MS}(m_c) = (1.27 \pm 0.02)\, \text{GeV}$ we find using three-loop running with $\alpha_s(m_c) = 0.395$ and $\alpha_s(m_Z) = 0.1185$ down to the hard-collinear scale $m_c^{\rm MS}(1.5\, {\rm GeV}) = 1.19\,{\rm GeV}$ {as central value at $1.5\, {\rm GeV}$}.
The change of  the 
hard-collinear scale indicated above then leads to $1.14\, {\rm GeV} \leq m_c \leq 1.26\, {\rm GeV}$. The parametric errors of ${m}_c^{\rm MS}(m_c)$ and $\alpha_s$ are neglected in view of the larger uncertainty due to the change of the hard-collinear scale $\mu_{\rm hc}$. 
{In contrast, two-loop running was used in the recent analysis in Ref.~\cite{Gunawardana:2019gep}, which 
gives  the value $m_c^{\rm MS}(1.5\, {\rm GeV}) = (1.20 + 0.03)\,{\rm GeV}$. Taking into account the parametric uncertainties, but no change of the hard-collinear scale, finally leads to the variation of the charm mass, $1.17\, {\rm GeV} \leq m_c \leq 1.23\, {\rm GeV}$, which was used in the analysis in Ref.~\cite{Gunawardana:2019gep}. 
As will be shown later, the different variation of the charm mass parameter in our present analysis compared to the one used  in the recent analysis in Ref.~\cite{Gunawardana:2019gep} turns out to be one of the main reasons for the discrepancy between the two analyses. } 

We note that  in the original analysis in Ref.~\cite{Benzke:2010js} just  $m_c(1.5\, {\rm GeV}) = 1.131\, {\rm GeV}$ was used and uncertainties were neglected. 
{ As already emphasized by the authors of Ref.~\cite{Gunawardana:2019gep},
controlling the scale dependence by calculating $\alpha_s$ corrections {to the resolved contributions}  would also help to better control the uncertainty due to the charm quark mass.} 

\newpage

For the operator basis 
we refer the reader to the original analysis in Ref.~\cite{Benzke:2017woq}. We calculate the uncertainty due to the resolved contributions relative to the decay rate in the OPE region.\footnote{For the $\bar B \to X_{s,d} \ell^+\ell^-$ case this means that there is no cut in the  hadronic mass and for the 
$\bar B \to X_s \gamma$ case the cut on the photon region is taken at a value around $E^{\rm cut}_\gamma  = 1.6\mbox{ GeV}$. We use the NLO OPE result of  the $\bar B \to X_{s,d} \ell^+\ell^-$ decay rate as in the original analysis in Ref.~\cite{Benzke:2017woq} and the LO one of the  $\bar B \to X_s \gamma$ rate as in the original analysis in  Ref.~\cite{Benzke:2010js}.} Therefore, the Wilson coefficients of the OPE result are naturally calculated at the hard scale. 

{ The Wilson coefficients in the resolved contribution are taken at the hard scale but at leading accuracy because we do not consider any $\alpha_s$ corrections or any RG improvements in the calculation of the resolved power corrections. 
{In this analysis} we then vary  the scale of the Wilson coefficients  in the resolved contributions between the hard and the hard-collinear scale {  -- while keeping the hard scale in the OPE rate fixed --} to make the scale dependence  of the results manifest.}~\footnote{ In the original~\cite{Benzke:2010js} and also in the recent 
analysis~\cite{Gunawardana:2019gep} the authors have chosen the hard-collinear scale for the Wilson
coefficient in the OPE rate which is {\it not} the natural scale of the OPE rate, in spite of the fact that the OPE rate at higher orders  is often  calculated at a scale slightly smaller  than the hard scale for other reasons (see i.e. Ref.~\cite{Misiak:2015xwa}).   For the Wilson coefficients in the resolved contribution these authors again  use the hard-collinear scale.  We note that using the hard or the hard-collinear scale in both, in the OPE rate and in the resolved contribution, leads only to a relatively small change of the final  result. The real scale ambiguity of the final result is  explored in the present analysis when we keep the hard scale in the OPE rate and vary the scale in resolved contribution from the hard to  the hard-collinear scale.}

In this work we mainly consider the resolved contribution due to the interference ${\cal O}^{c}_{1} - {\cal O}_{7\gamma}$, which is the numerically most relevant for the case $\bar B \to X_{s ,d} \ell^+ \ell^-$, but also for the case $\bar B \to X_{s} \gamma$.
The explicit form of the subleading shape function for that contribution was derived in Ref.~\cite{Benzke:2010js}:
\begin{equation}
h_{17}(\omega_1,\mu) = \int\frac{dr}{2\pi}\,e^{-i\omega_1r}\frac{\braB \bar{h} (0) \nbs i \gamma_\alpha^\perp\nb_\beta g  G^{\alpha\beta}(r\nb)h(0)\ketB}{2M_B}\,,
\end{equation}
where $n$ and $\nb$ are the light-cone vectors and $h$ and $G$ are the heavy quark and gluon field, respectively. Soft Wilson lines connect the fields to ensure gauge invariance but are suppressed in the notation.
The variable $\omega_1$ corresponds to the soft gluon momentum. (The integration over $\omega$ which is related to the heavy quark momentum is already taken here.)

With the help of standard HQET techniques one can derive from PT invariance that the function $h_{17}$ is real and even in $\omega_1$. 
The new {estimates of}  the moments of this subleading shape function in the interference term ${\cal O}^{c}_{1} - {\cal O}_{7\gamma}$ as derived in Ref.~\cite{Gunawardana:2019gep} lead to the additional constraints
\begin{eqnarray}
\int^\infty_{-\infty} d\omega_1 \, {\omega_1}^0 \, h_{17} (\omega_1,\mu) &= (0.237 &\pm\, 0.040)\, {\rm GeV^2}\,, \nonumber \\
\int^\infty_{-\infty} d\omega_1 \, {\omega_1}^2 \, h_{17} (\omega_1,\mu) &= (0.15 &\pm\, 0.12)\, {\rm GeV^4}\,. 
\label{moments}
\end{eqnarray}
The normalisation was already known before. The second moment has been used for the first time in the case of $\bar B \to X_s \gamma$ in Ref.~\cite{Gunawardana:2019gep}. All odd moments of $h_{17}$ in $\omega_1$ vanish because the function is even. It is worth noting that more moments can be expressed in terms 
of HQET parameters as was shown in Refs.~\cite{Gunawardana:2017zix,Gunawardana:2019gep}, thus more accurate determinations of the moments might be possible in the future. 

{However, we note that the determination of the HQET parameters related 
to the second and also higher moments are based on the so-called Lowest-Lying State Approximation (LLSA) 
(see Refs.~\cite{Mannel:2010wj,Heinonen:2014dxa,Heinonen:2016cwm}). {This method allows to estimate  
higher-dimensional operators (related to the higher moments) by assuming that the lowest lying heavy meson state saturate a sum-rule for the insertion of a heavy meson state sum. This way LLSA relates higher-dimensional matrix elements to the known lower-dimensional ones. In Ref.~\cite{Gambino:2016jkc} the error due to this approximation was estimated to be $60-100\%$. This large uncertainty also enters the second equation in Eq.~\ref{moments}.}

{The natural scale of the HQET parameters related to the second moment is of  $O( \Lambda^4_{\rm QCD} ) $ or even higher powers of $\Lambda_{\rm QCD}$ in case of the parameters related to higher moments. This in principle allows for a rough dimensional analysis of the n-th moment to be a linear combination of parameters of 
order $\Lambda_{\rm QCD}^{n+2}$ with O(1) coefficients, a feature which is confirmed 
in existing HQET calculations, in particular in the case of the second moment of $h_{17}$. Also the fourth and the sixth moment can be expressed by parameters of $\Lambda_{\rm QCD}^6$ and $\Lambda_{\rm QCD}^8$, respectively.  {\it Assuming} that the coefficients are still of O(1) or only slightly larger in case of the sixth moment one gets led
to the following dimensional estimates
\begin{eqnarray}
-0.3\,{\rm GeV^6}\, \lesssim \,  \int^\infty_{-\infty} d\omega_1 \, {\omega_1}^4 \, h_{17} (\omega_1,\mu) &\, \lesssim   \,  + 0.3\,{\rm GeV^6}\,, \nonumber \\
-0.3\,{\rm GeV^8}\, \lesssim  \,  \int^\infty_{-\infty} d\omega_1 \, {\omega_1}^6 \, h_{17} (\omega_1,\mu) &\,  \lesssim  \,  + 0.3\,{\rm GeV^8}\,.
\label{highermoments}
\end{eqnarray}
These estimates were also used in a similar way in the analysis in Ref.~\cite{Gunawardana:2019gep}; we consistently use these estimates   for all model functions within the present analysis.}~\footnote{However, we note that  to our knowledge there is no general argument that for the unknown higher moments the coefficients of HQET parameters scaling with a certain power of $\Lambda_{\rm QCD}$ are always O(1). A counter example is given  by the  model function for the subleading shape function $h_{17} = exp(-| x/\Lambda |)$
for which we find $\int^\infty_{-\infty} d\omega_1\,{\omega_1}^n\,exp(-| x/\Lambda |)  = \Lambda\, ((-\Lambda)^n + \Lambda^n)\,\Gamma(1+n)$. Here the second moment is of order $\Lambda^3$  with a coefficient  4, the fourth moment is of order 
$\Lambda^5$  with a  coefficient 48 and the sixth moment  is of order $\Lambda^7$  with a coefficient 1440 (!).  Therefore, we analyse  the impact of these two additional dimensional estimates within our analysis,
and this way we offer the results to the reader also for the case when no such estimates on the higher moments are used. }

Finally, one assumes that the subleading shape function as a soft function should not have any significant structures like maxima outside the hadronic range ($-1\,$GeV$<\omega_1<1\,$GeV) and the values of it should be within the hadronic range ($-1\,$GeV$<h_{17}(\omega_1)<1\,$GeV).
In the following we will take all those properties into account when we consider model functions in the convolution with the jet function.

{Nothing further is known about the structure of the 
subleading shape functions. Thus, we follow the strategy used by authors of Ref.~\cite{Gunawardana:2019gep} 
who modelled the shape function $h_{17}$ by using a complete set of basis functions. This systematic approach was already advocated before and used in several applications~\cite{Ligeti:2008ac,Lee:2008xc,Bernlochner:2020jlt}.} 
{In the original analyses in Refs.~\cite{Benzke:2010js,Benzke:2017woq}  simple functions like polynomials of second degree multiplied by a Gaussian function were used. The systematic approach  using a complete basis of model functions allows to avoid any prejudice  regrading the functional form of the shape functions.}

{Due to the importance of the resolved ${\cal O}^{c}_{1} - {\cal O}_{7\gamma}$  contribution for the overall uncertainty in the decay $\bar B \to X_s \gamma$ we first revisit the recent  analysis in Ref.~\cite{Gunawardana:2019gep} in Section 2. 
We will extend our findings 
to decay $\bar B \to X_{s ,d} \ell^+ \ell^-$ in Section 3. Section 4 is reserved for our summary and our conclusions.}

\section{Resolved contributions   
to the decay $\bar B \to X_s \gamma$}

The relative uncertainty of the decay rate of $\bar B \to X_s \gamma$ due to the non-local resolved contribution within the interference 
of ${\cal O}_1- {\cal O}_{7\gamma}$~\footnote{To simplify the notation we leave out the  superscript  "c"  in the following.} is given by
\begin{equation}\label{relative uncertainty}
  {\cal F}_{\rm b \to s \gamma}^{17} = \frac{C_1(\mu)\, C_{7\gamma}(\mu)}{(C_{7\gamma}(\mu_{\rm \mbox{{\tiny OPE}}}))^2}\, \frac{\Lambda_{17}(m_c^2/m_b,\mu)}{m_b}\,, 
\end{equation}
where at order $1/m_b$ one finds~\cite{Benzke:2010js}:
\begin{equation}\label{Lambda17A}
  \Lambda_{17}\Big(\frac{m_c^2}{m_b},\mu\Big)
  = e_c\,\mbox{Re} \int_{-\infty}^\infty \frac{d\omega_1}{\omega_1} 
  \left[ 1 - F\!\left( \frac{m_c^2-i\varepsilon}{m_b\,\omega_1} \right)
  + \frac{m_b\,\omega_1}{12m_c^2} \right] h_{17}(\omega_1,\mu)\,,
\end{equation}
with the penguin function $F(x) = 4\, x\, {\rm arctan}^2(1/\sqrt{4x-1})$.

We start with the model function used in the original analyses in Refs.~\cite{Benzke:2010js,Benzke:2017woq}, namely a polynomial of second degree
combined with a Gaussian function: 
\begin{equation}
h_{17}(\omega_1)=\frac{2\lambda_2}{\sqrt{2\pi}\sigma}\frac{\omega_1^2-\Lambda^2}{\sigma^2-\Lambda^2}e^{-\frac{\omega_1^2}{2\sigma^2}}\,,
\label{eqn:h17b}
\end{equation}
in which the two hadronic parameters, $\Lambda$ and $\sigma$, are chosen to be of order $\Lambda_{\text QCD}$. 
Combining this function with all constraints mentioned in the last section, one finds that the reduction of the uncertainty due to the resolved contributions in the decay $\bar B \to X_s \gamma$ is two-fold:
\begin{itemize}
\item First, the central value of the charm mass at the hard-collinear scale moved from $m_c(1.5\,{\rm GeV}) = 1.131\,{\rm GeV}$ 
used in the original analysis in Ref.~\cite{Benzke:2010js} to $m_c(1.5\,{\rm GeV}) = 1.19\,{\rm GeV}$ in the recent analysis in 
Ref.~\cite{Gunawardana:2019gep}, and the central value of the bottom mass in the shape function scheme moved from $m_b=4.65\,\text{GeV}$ to the new value $m_b=4.58\,\text{GeV}$. 
As shown in the upper plot of Fig.\ref{Fig:bsgamma}, these changes in the input parameters have the effect that the jet function moves slightly outside the hadronic range and the overlap and therefore the convolution integral with the model function becomes smaller. {The dependence on  the charm mass is pronounced.} 
{Varying the charm mass will therefore have a noticeable impact on the resolved contribution,  
 leading to larger values than in the recent  analysis in Ref.~\cite{Gunawardana:2019gep}.}

\item Second, the new bound on the second moment of the shape function, given in Eq.~\ref{moments}, significantly restricts the shape of the soft function and consequently leads to a reduction of the extreme values of the convolution integral as shown in the bottom plot of Fig.\ref{Fig:bsgamma}. 
\end{itemize}

\begin{figure}[hpt]
  \centering\includegraphics[scale=1.0]{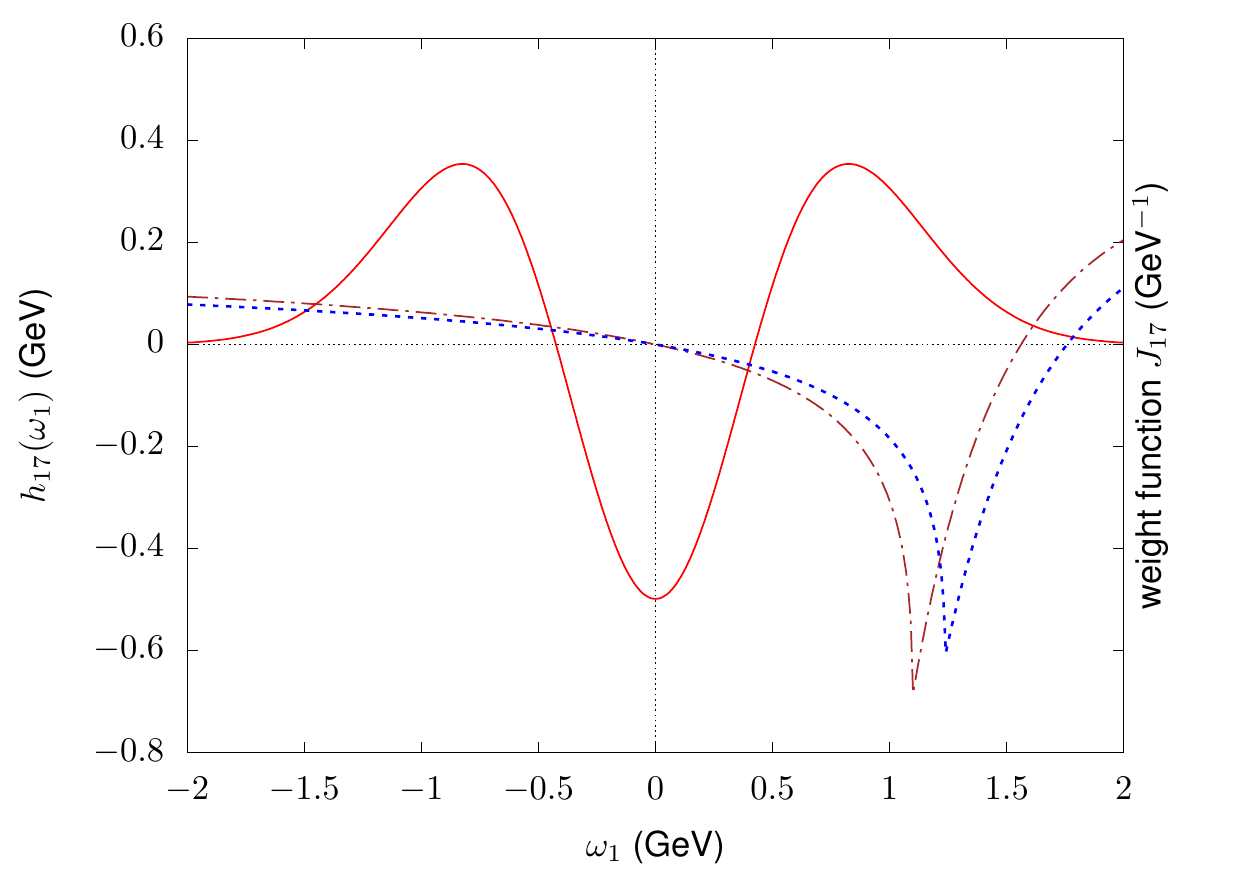} \includegraphics[scale=1.0]{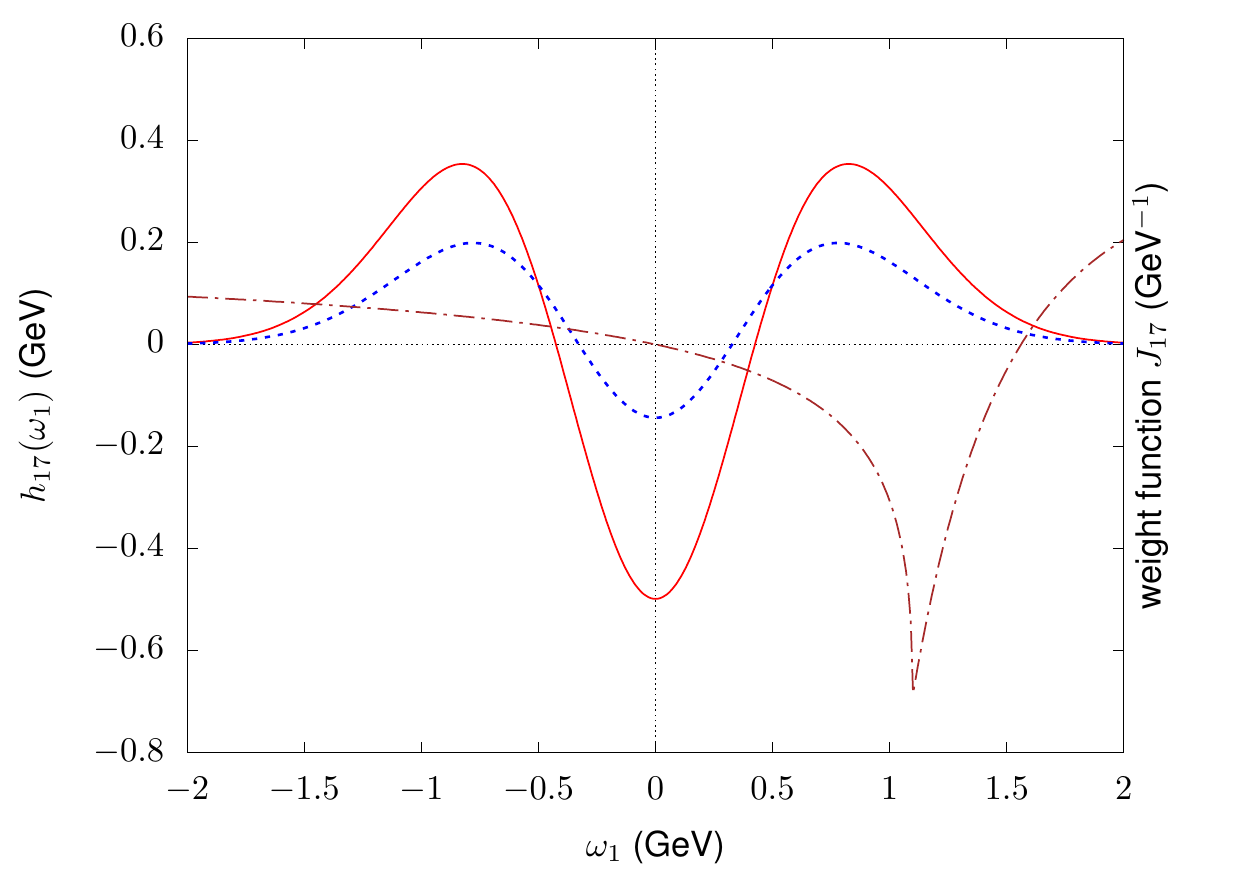}
   \caption{The top figure shows the jet (weight) function in the case $\bar B \to X_s \gamma$ for $m_c= 1.131$\,GeV and $m_b = 4.65$\,GeV (dashed dotted, brown) 
  and for $m_c= 1.19$\,GeV and $m_b = 4.58$\,GeV (dotted, blue) with the shape function in Eq.~\ref{eqn:h17b} (solid, red). The bottom figure shows in addition the shape function with a second moment which satisfies the new constraint (dotted, blue). \label{Fig:bsgamma}}
\end{figure}

In the recent analysis~\cite{Gunawardana:2019gep} the authors modelled the shape function $h_{17}$ by using a complete set of basis functions, namely  the Hermite polynomials multiplied by a Gaussian\footnote{The Hermite polynomials are orthogonal with respect to a weight function $e^{-x^2}$, so that we have 
\begin{equation} 
\int^\infty_{-\infty} H_m(x) H_n(x) e^{-x^2} dx = \pi^{1/2} 2^n n!\, \delta_{nm}\,. \nonumber
\end{equation} 
The Hermite polynomials form an orthogonal basis of the Hilbert space of functions which satify
$\int^\infty_{-\infty} |f(x)|^2 e^{-x^2} dx < \infty .$
The inner product is defined as $\langle f , g \rangle = \int^\infty_{-\infty} f(x) \overline{g(x)} e^ {-x^2} dx$.}\, in order to make a systematic analysis of all possible 
model functions { - as already advocated by the authors of Ref.~\cite{Ligeti:2008ac}.} 
{This systematic approach allows to avoid any prejudice  regrading the unknown functional form of the shape functions. We note here that in the original analyses in Refs.~\cite{Benzke:2010js,Benzke:2017woq}
simple functions like a second-order polynomial with a Gaussian function were assumed.}

Because the shape function $h_{17}$ is even, one needs only even polynomials in the systematic expansion:
\begin{equation}\label{Hermite} 
h_{17}(\omega_1)=\sum_n a_{2n} H_{2n}\left(\frac{\omega_1}{\sqrt{2}\sigma}\right)e^{-\frac{\omega_1^2}{2\sigma^2}}. 
\end{equation}
The Hermite polynomials are very suitable for this purpose because they are orthogonal and, thus, the $2k$-th moment of $h_{17}$ only depends on the 
coefficients $a_{2n}$ with $n \leq k$. Therefore, the zeroth moment only depends on $a_0$ and the second moment depends on $a_0$ and $a_2$. 
This also means that the first $2k$ moments determine $a_{2n}$ with $n \leq k$~\cite{Gunawardana:2019gep}.

{Our present analysis follows the strategy of Ref.~\cite{Gunawardana:2019gep}, but we rigorously explore the space of Hermite polynomials multiplied by  a Gaussian. Hermite polynomials with ${\rm exp}(-x^4)$ or ${\rm exp}(-x^6)$ suppression can also be expressed in the basis above, but}  this requires an infinite sum and is therefore not considered in an approach that only takes into account a limited number of terms. The recent analysis~\cite{Gunawardana:2019gep} does not consider polynomials with a degree higher than $10$.}  {We  anticipate} that the extreme values for the uncertainty are realised with {polynomials of degree $6$ with 
an ${\rm exp}(-x^2)$ suppression  or with polynomials of degree $4$ and $6$} with an ${\rm exp}(-x^4)$ suppression and that  {already} polynomials of degree $8$ and higher suppression factors like ${\rm exp}(-x^6)$ do not lead to larger values.

Our grid of input parameters of the model function is the following: We scan through the one-sigma ranges of the input parameters 
$1.14\,\text{GeV} \leq m_c \leq 1.23\,\text{GeV}$ with 10 steps, $4.55\,\text{GeV} \leq m_b \leq 4.61\,\text{GeV}$ with 3 steps, the first moment $m_0$ from $0.197\,{\rm GeV^2}$ to $0.277\,{\rm GeV^2}$ with 8 steps and the second moment $m_2$ from $0.03\,{\rm GeV^4}$ 
to $0.27\,{\rm GeV^4}$ with 12 steps, {and also the fourth and the sixth moment between 
$-0.3\,{\rm GeV^6}$ and $0.3\,{\rm GeV^6}$ and between $-0.3\,{\rm GeV^8}$ and $0.3\,{\rm GeV^8}$, respectively, in 30 steps.} Moreover, we vary the hadronic parameter $\sigma$ from $-1\, \text{GeV}$ to $+1\, \text{GeV}$ in 40 steps.  

We also  {anticipate} that -- except for the upper bound in case of the sum of Hermite polynomial of degree 0 and 2 -- the extreme values of $\Lambda_{17}$ for all the different model functions can be found using the mass parameters $m_c = 1.14\,{\rm GeV} $ and $m_b = 4.61\,{\rm GeV}$. This is expected, since for any larger value of $m_c$ and any smaller value of $m_b$ the jet function moves further out of the hadronic range (see Fig.~\ref{Fig:bsgamma}).

{In the case of the model function with the sum of $n=0$ and $n=2$ polynomials (see Eq.~\ref{Hermite}) we find in our multi-parameter scan 
\begin{equation}
- 24\,\text{MeV} \leq \Lambda_{17} \leq - 1\,\text{MeV}\hspace{1cm}(n \leq 2,{\rm exp}(-x^2)).
\end{equation} 
The lower bound is found with $\sigma = 400\,\text{MeV}$, with the zeroth moment 
$m_0 = 0.200\,\text{GeV}^2$ and with the second moment $m_2 = 270\,\text{GeV}^4$. This implies for the higher moments $m_4 = 0.244\,\text{GeV}^6$
and $m_6 = 0.286\,\text{GeV}^8$. The upper bound corresponds to the parameter set, $\sigma = 140\, \text{MeV}$, $m_0 = 0.280\,\text{GeV}^2$, and $m_2 = 0.0030\,\text{GeV}^4$. The sum of $n=0$, $n=2$, and $n=4$ polynomials leads to 
 \begin{equation}
- 27  \,\text{MeV} \leq \Lambda_{17} \leq + 4\,\text{MeV}\hspace{1cm}(n \leq 4,{\rm exp}(-x^2)).
\end{equation} 
The lower bound corresponds to the parameter set $\sigma = 300\, \text{MeV}$, $m_0 = 0.260\,\text{GeV}^2$, $m_2 = 0.270\,\text{GeV}^4$, 
 and $m_4 = 0.260\,\text{GeV}^6$, the upper bound to $\sigma = 340\, \text{MeV}$, $m_0 = 0.220\,\text{GeV}^2$, $m_2 = 0.030\,\text{GeV}^4$, and 
$m_4 = - 0.100\,\text{GeV}^6$. An even larger interval is found with a sum of Hermite polynomials up to order 6, namely
 \begin{equation}
- 29  \text{MeV} \leq \Lambda_{17} \leq + 6\,\text{MeV}\hspace{1cm}(n \leq 6,{\rm exp}(-x^2)),
\end{equation} 
with the lower bound corresponding to the parameters $\sigma = 280\, \text{MeV}$, $m_0 = 0.200\,\text{GeV}^2$, $m_2 = 0.270\,\text{GeV}^4$,
$m_4 = 0.280\,\text{GeV}^6$, and $m_6 = 0.300\,\text{GeV}^8$ and the upper bound with $\sigma = 300\, \text{MeV}$, $m_0 = 0.200\,\text{GeV}^2$, $m_2 = 0.030\,\text{GeV}^4$, $m_4 = - 0.120\,\text{GeV}^6$, and $m_6 = - 0.220\,\text{GeV}^8$.\\ With an additional polynomial of degree 8 one does not find larger values:\footnote{We note that in contrast to the authors of the recent paper~\cite{Gunawardana:2019gep} we also find solutions with polynomials up to degree 8 due to our more dense grid.} 
 \begin{equation} 
- 29 \,\text{MeV} \leq \Lambda_{17} \leq + 6\,\text{MeV}\hspace{1cm}(n \leq 8,{\rm exp}(-x^2)).
\end{equation} 
The lower bound is obtained for $\sigma = 260\, \text{MeV}$, $m_0 = 0.280\,\text{GeV}^2$, $m_2 = 0.270\,\text{GeV}^4$,
$m_4 = 0.260\,\text{GeV}^6$, $m_6 = 0.300\,\text{GeV}^8$, and $m_8 = 0.380\,\text{GeV}^{10}$, the upper bound for $\sigma = 300\, \text{MeV}$, $m_0 = 0.280\,\text{GeV}^2$, $m_2 = 0.030\,\text{GeV}^4$, $m_4 = - 0.120\,\text{GeV}^6$, $m_6 = - 0.220\,\text{GeV}^8$, and $m_8 = - 0.340\,\text{GeV}^{10}$.

If one uses model functions with ${\rm exp}(-x^4)$ or ${\rm exp}(-x^6)$ suppression instead of a Gaussian (${\rm exp}(-x^2)$) one still finds {slightly} larger intervals for $\Lambda_{17}$. In case of the Hermite polynomials up to degree 4 with a weight function $exp(-x^4)$ one gets 
 \begin{equation}
- 31 \,\text{MeV} \leq \Lambda_{17} \leq + 9\,\text{MeV}\hspace{1cm}(n \leq 4,{\rm exp}(-x^4)).\label{Maximum}
\end{equation} 
The lower bound corresponds to the parameter set $\sigma = 740\, \text{MeV}$, $m_0 = 0.280\,\text{GeV}^2$, $m_2 = 0.270\,\text{GeV}^4$, $m_4 = 0.300\,\text{GeV}^6$ and the upper bound to $\sigma = 800\, \text{MeV}$, $m_0 = 0.200\,\text{GeV}^2$, and $m_2 = 0.030\,\text{GeV}^4$ and $m_4 = - 0.120\,\text{GeV}^6$. 
With the Hermite polynomials up to degree 6 with an ${\rm exp}(-x^4)$ suppression, one obtains the same result:
 \begin{equation}
- 31\,\text{MeV} \leq \Lambda_{17} \leq + 9\,\text{MeV}\hspace{1cm}(n \leq 6,{\rm exp}(-x^4)). \label{Maximumb}
\end{equation} 
The corresponding parameter sets are $\sigma = 720\, \text{MeV}$, $m_0 = 0.200\,\text{GeV}^2$, $m_2 = 0.270\,\text{GeV}^4$,
$m_4 = 0.440\,\text{GeV}^6$, and $m_6 = 0.580\,\text{GeV}^8$ for the lower bound and $\sigma = 760\, \text{MeV}$, $m_0 = 0.280\,\text{GeV}^2$, $m_2 = 0.030\,\text{GeV}^4$, $m_4 = - 0.120\,\text{GeV}^6$, and $m_6 = - 0.200\,\text{GeV}^8$ for the upper bound.
If one uses a higher suppression, namely ${\rm exp}(-x^6)$ for example with a Hermite polynomial up to degree 4, one gets a significantly smaller interval, namely 
 \begin{equation}
- 29  \,\text{MeV} \leq \Lambda_{17} \leq +1\,\text{MeV}\hspace{1cm}(n \leq 4,{\rm exp}(-x^6)),
\end{equation} 
with $\sigma = 900\, \text{MeV}$, $m_0 = 0.200\,\text{GeV}^2$, $m_2 = 0.270\,\text{GeV}^4$, $m_4 = - 0.300\,\text{GeV}^6$ for the lower bound and to $\sigma = 900\, \text{MeV}$, $m_0 = 0.280\,\text{GeV}^2$, and $m_2 = 0.030\,\text{GeV}^4$ and $m_4 = 0.300\,\text{GeV}^6$ for the upper bound. 

{Summing up, the largest interval we find is $ - 31\,\text{MeV} \leq \Lambda_{17} \leq + 9\,\text{MeV}$. Our new result has a $42\%$ smaller range than the original one in Ref.~\cite{Benzke:2010js}, $- 42\,\text{MeV} \leq \Lambda_{17} \leq + 27\,\text{MeV}\, $  where the model given in Eq.~\ref{eqn:h17b} and no constraint on the second, fourth and sixth moments was used. 
In the recent analysis in Ref.~\cite{Gunawardana:2019gep} a stronger  reduction by almost  $60\%$ compared to the result in Ref.~\cite{Benzke:2010js} was found, namely 
$ - 24\,\text{MeV} \leq \Lambda_{17} \leq + 5\,\text{MeV}\,$~\footnote{We note here that we have fully reproduced these results using their input and their assumption with our numerics.} The reasons for this discrepancy between our and the recent analysis in Ref.~\cite{Gunawardana:2019gep} are threefold: 
\begin{itemize}
\item {The important difference is the fact that we take into account a larger uncertainty due to the charm mass
as discussed in the introduction.} 
\item We use a denser grid of parameters to find the extrema of the resolved contributions. 
\item We use the fact that also polynomials with suppression factors ${\rm exp}(-x^4) $ or ${\rm exp}(-x^6)$ can be expressed in terms of the original basis given in Eq.~\ref{Hermite}, and, thus, have also to be considered within a systematic analysis. 
\end{itemize}
\vspace{+0.2cm} 
  }

A further subtlety arises from kinematic corrections.
The original analysis of the $\bar B \to X_s \gamma$ case included an additional large $1/m_b^2$ correction due to kinematic factors ~\cite{Benzke:2010js}. In order to make this manifest, Eq.~\ref{Lambda17A} should be replaced by
\begin{equation}\label{Lambda17B}
\begin{aligned}
  \Lambda_{17}\Big(\frac{m_c^2}{m_b},\mu\Big)
  &= e_c\,\mbox{Re} \int_{-\infty}^{\bar\Lambda}\!d\omega
  \int_{-\infty}^\infty \frac{d\omega_1}{\omega_1} \\
  &\quad\times \left\{ \left( \frac{m_b+\omega}{m_b} \right)^3 
  \left[ 1 - F\!\left( \frac{m_c^2-i\varepsilon}{(m_b+\omega)\,\omega_1} \right) \right]
  + \frac{m_b\,\omega_1}{12m_c^2} \right\} g_{17}(\omega,\omega_1,\mu) \,, 
\end{aligned}
\end{equation}
where  $h_{17}(\omega_1,\mu) = \int d \omega  \, g_{17}(\omega,\omega_1,\mu)$.\footnote{For the precise limits of integration we refer the reader to the discussion in Section 6 of Ref.~\cite{Benzke:2010js}.} Obviously, the factor $(m_b + \omega)$ was approximated by $m_b$ within the prefactor and within the function $F$ in Eq.~\ref{Lambda17A} at order $1/m_b$. If we include this $1/m_b^2$ effect, we find the extreme range for $\Lambda_{17}$ for the same parameters as in the cases without the $1/m_b^2$ correction. 
If one chooses a Gaussian suppression, it is again the sum of Hermitian polynomials up to degree 6 which leads to the largest interval: 
 \begin{equation}
- 54  \,\text{MeV} \leq \Lambda_{17} \leq - 1\,\text{MeV}\,.
\end{equation} 
And if one chooses a ${\rm exp}(x^{-4})$ suppression, the polynomials up to degree 4 and 6 lead again to the  maximal results:
 \begin{equation}
- 59 \,\text{MeV} \leq \Lambda_{17} \leq + 4\,\text{MeV}\,,
\end{equation} 
 \begin{equation}
- 61  \,\text{MeV} \leq \Lambda_{17} \leq + 5\,\text{MeV}\,. \label{Maximum2}
\end{equation} 
{This should be compared to $- 60\,\text{MeV} \leq \Lambda_{17} \leq + 25.0\,\text{MeV}$ found in the original analysis~\cite{Benzke:2010js}.  Our final result shows  a reduction of the uncertainty of approximately $25\%$.

We emphasise  that this $1/m_b^2$ piece directly originates from the ${\cal O}_1 - {\cal O}_{7\gamma}$ contribution as shown above.  It has a large numerical impact increasing this resolved contribution by {more than $50\%$}.
In contrast, resolved contributions like the ones  due to  the operator pairs ${\cal O}_1 -  {\cal O}_{8 g}$  or ${\cal O}_1 - {\cal O}_{1}$   which also occur at the order $1/m_b^2$ were shown to be numerically negligible in the original analysis~\cite{Benzke:2010js}.
The recent analysis in Ref.~\cite{Gunawardana:2019gep} did not take this $1/m_b^2$ correction into account.
\begin{itemize}
\item Thus, dropping this numerically large $1/m_b^2$ term {represents a large piece of reduction of the uncertainty in the analysis in Ref.~\cite{Gunawardana:2019gep}       compared to the original analysis  in Ref.~\cite{Benzke:2010js} and also represents the second important  difference to our present analysis.}
\end{itemize}
\vspace{0.2cm}

{Finally, we analyze the impact of the dimensionally estimated bounds on the fourth and the sixth moment
given in Eqs.~(\ref{highermoments}). Without these estimates we  would find the extreme values again for the Hermite 
polynomials up to degree 4 or 6 with a suppression factor ${\rm exp}(-x^4)$, namely 
$- 72 \,\text{MeV} \leq \Lambda_{17} \leq + 4\,\text{MeV}$ and  $- 76  \,\text{MeV} \leq \Lambda_{17} \leq + 5 \,\text{MeV}$. But also with polynomials up to degree 6 and a Gaussian suppression we would already get a rather large result: $- 63\,\text{MeV} \leq \Lambda_{17} \leq + 1\,\text{MeV}$.
The direct comparison of these results with the extreme one we have found using the dimensionally estimated  bounds given in Eqs.(\ref{highermoments}), shows their  large impact.}\\

{\bf Summary of numerical results in the case of $\bar B \to X_s \gamma$:} Our result for $\Lambda_{17}$ at order $1/m_b$, $ - 31\,\text{MeV} \leq \Lambda_{17} \leq + 9\,\text{MeV}$\,, as given in Eqs.~(\ref{Maximum}) and (\ref{Maximumb}),  translates into the following relative uncertainty of the decay rate of $\bar B \rightarrow X_s \gamma$ via Eq.~\ref{relative uncertainty}:
\begin{equation}
{\cal F}_{\rm b \to s \gamma}^{17} |_{1/m_b} \in [-0.7\%,\,2.4\%]\,,
\end{equation}
which is significantly larger than the result of the recent analysis in Ref.~\cite{Gunawardana:2019gep}. 
but also significantly smaller  than  the  corresponding result in the original analysis in Ref.~\cite{Benzke:2010js}. Several  reasons for this difference to the result in  Ref.~\cite{Gunawardana:2019gep}  were indicated in detail in our analysis.
{The most important one is that we use a larger uncertainty in the charm mass  {(as discussed in the introduction)}  compared to the analysis in 
Ref.~\cite{Gunawardana:2019gep}. }

If we include the large additional $1/m_b^2$ piece - as {\it not} done in the recent analysis in Ref.~\cite{Gunawardana:2019gep}  -  our result, $ - 61,\text{MeV} \leq \Lambda_{17} \leq + 5\,\text{MeV}\, $, 
as given in Eq.~\ref{Maximum2}, leads to our final result:
\begin{equation}
 {\cal F}_{\rm b \to s \gamma}^{17} \in  [-0.4\%,\,4.7\%]\,,  
\end{equation}
{It was shown in \cite{Benzke:2010js} that this kinematical $1/m_b^2$ contribution from the ${\cal O}_1 - {\cal O}_{7\gamma}$ interference  is the only numerically relevant contribution  at the second order in $1/m_b$.}
Our result represents a significant reduction of the uncertainty compared to the result of the original analysis in Ref.~\cite{Benzke:2010js}, ${\cal F}_{\rm b \to s \gamma}^{17} \in [-1.9\%,\,4.7\%] $, but is still much larger than the result in the recent analysis in Ref.~\cite{Gunawardana:2019gep},  ${\cal F}_{\rm b \to s \gamma}^{17} \in [-0.4\%,\,1.9\%] $ {which is missing the large $1/m_b^2$ contribution}.   {These latter numbers of  Ref.~\cite{Benzke:2010js} and  of  Ref.~\cite{Gunawardana:2019gep}  are translated to  our scale fixing.}\footnote{The numbers do not agree with the quoted ones in the original analysis Ref.~\cite{Benzke:2010js}
because the authors use the  hard-collinear scale in the Wilson coefficients of  the resolved contribution and also in the Wilson coefficients of  the OPE rate. The same scale fixing was used in the recent analysis Ref.~\cite{Gunawardana:2019gep}. In contrast, we have chosen  the hard scale as our default value within the resolved contribution as mentioned in the introduction and the OPE rate is naturally fixed at the hard scale. {Using their scale-fixing (with the OPE rate and the resolved contribution fixed at the hard-collinear scale)  one finds    ${\cal F}_{\rm b \to s \gamma}^{17} \in [-1.7\%,\,4.0\%]$ in the original analysis in Ref.~\cite{Benzke:2010js} and ${\cal F}_{\rm b \to s \gamma}^{17} \in [-0.3\%,\,1.6\%]$ in the recent  analysis in  Ref.~\cite{Gunawardana:2019gep}.}}

 {If we do} not use the dimensional estimates on the higher moments, given in Eq. (\ref{highermoments}),  { we  find} a much larger uncertainty, ${\cal F}_{\rm b \to s \gamma}^{17} |_{1/m_b} \in [-0.4\%,\,5.9\%]$ what shows the  large impact of these dimensional estimates.} 

{Finally, we consider scale variations in our final result.  The present results are leading order results, no $\alpha_s$ corrections are calculated and no 
RG improvements were implemented. The only scale in our resolved contribution is within the hard function, represented by the Wilson coefficients. 
Therefore we have chosen the scale in the Wilson coefficients of the resolved contribution at the hard scale as our 
default value. If we run down the LO Wilson coefficients  $C_1(\mu)\, C_{7\gamma}(\mu)$ to the hard-collinear scale {and keep the OPE rate at the hard scale}, the result increases by more than  $40\%$ compared to our default value. There is no strict argument here that this specific scale variation in our result  can be connected to an estimate of the unknown NLO corrections. However,  this observation calls for a calculation of the $\alpha_s$ corrections  and RG resummations.}

We also emphasize that the local Voloshin term\footnote{This local term can be derived from the resolved contribution ${\cal O}^{c}_{1} - {\cal O}_{7\gamma}$  by neglecting the shape function effects and under the assumption that the charm quark mass is treated as heavy 
(see section 3.2 of Ref.~\cite{Benzke:2017woq}). It was shown that this local term derived in 
Refs.\cite{Voloshin:1996gw,Ligeti:1997tc,Grant:1997ec,Buchalla:1997ky} does not fully account for the corresponding resolved contribution.} is subtracted from the resolved contribution  ${\cal F}_{\rm b \to s \gamma}^{17}$. This has been traditionally done in all analyses of this specific resolved contribution to the $\bar B \to X_s \gamma$ decay rate. Therefore this local Voloshin term
has still to be added to the decay rate. It corresponds to $\Lambda_{17}^{\rm Voloshin} = (-1) (m_b \lambda_2)/(9 m_c^2)$ 
which translates in
\begin{equation} 
{\cal F}_{\rm b \to s \gamma}^{\rm Voloshin} = - \frac{C_1\,C_{7\gamma}\, \lambda_2}{(C_{7\gamma})^2\, 9\, m_c^2} = +3.3\%\,, 
\end{equation}

There are two more resolved contributions at order $1/m_b$ as discussed in the introduction. In the original analysis in Ref.~\cite{Benzke:2010js} the resolved contributions due to the interference ${\cal O}_{7\gamma} - {\cal O}_{8g}$ and ${\cal O}_{8g} - {\cal O}_{8g}$ were estimated to 
$ {\cal F}_{\rm b \to s \gamma}^{78,{\rm VIA}} = [-3.0\%,\,-0.3\%] $  
and 
$ {\cal F}_{\rm b \to s \gamma}^{88} = [-0.3\%,\,2.1\%] $, using our scale fixing.   
The superscript ${\rm VIA}$ indicates that the resolved contribution ${\cal F}^{78}$ was determined by using the vacuum insertion approximation. 
We add up the three contributions using the scanning method and arrive at the final result for all resolved contributions: 
\begin{equation} 
{\cal F}_{\rm b \to s \gamma}^{\rm total} \in [-3.7\%,\,6.5\%]     \quad  {\rm (VIA)} . 
\end{equation}
{This has to be compared to the final result in the original analysis, which reads when translated to our default scales:  
${\cal F}_{\rm b \to s \gamma}^{\rm total} \in [-5.2\%,\,6.5\%] $}.

We finally note, that there is an alternative estimation of ${\cal F}^{78}$ offered in Ref.~\cite{Benzke:2010js} based on experimental data on $\Delta_{0-}$, the isospin asymmetry of inclusive neutral and charged $B \to X_s \gamma$ decay  using Babar measurements~\cite{Aubert:2005cua,Aubert:2007my}.   In the recent analysis~\cite{Gunawardana:2019gep}, the authors derived new bounds based on the inclusion of a new Belle measurement of $\Delta_{0-}$~\cite{Watanuki:2018xxg}, which leads to the experimental determination of ${\cal F}^{78}$  being the same order of magnitude as the determination using VIA.

\section{Resolved contributions  
to the decay $\bar B \to X_{s,d} \ell^+\ell^-$}

{We now update our analysis in Ref.~ \cite{Benzke:2017woq} using the {new estimate of} the second moment of the shape function $h_{17}$.}
In the case of the decay  $\bar B \to X_{s} \ell^+\ell^-$  the relative contribution due to the interference of ${\cal O}_1$ with ${\cal O}_{7\gamma}$ is given at order $1/m_b$ by 
\begin{align}
&{\mathcal F}^{17}_{\rm b \to s \ell \ell}
=\frac{1}{m_b}\frac{C_1(\mu)C_{7\gamma}(\mu)}{C_{\rm OPE}} e_c\,
\int_{-\infty}^{+\infty}d\omega_1\,
J_{17}(q_\mathrm{min}^2,q_\mathrm{max}^2,\omega_1)\,
h_{17}(\omega_1,\mu)\,,
\label{eqn:bsll}
\end{align}
where the shape function $h_{17}$ is the same one as in the decay $\bar B \to X_{s} \gamma$ and the jet function is given by 
\begin{align}
&J_{17}(q_\mathrm{min}^2,q_\mathrm{max}^2,\omega_1) = 
\mathrm{Re} \frac{1}{\omega_1+\ie}
\int_{\frac{q_\mathrm{min}^2}{M_B}}^{\frac{q_\mathrm{max}^2}{M_B}} \frac{d\nb\cdot q}{\nb\cdot q}\,
\frac{1}{\omega_1} \nn\\
&\left[
(\nb\cdot q+\omega_1)\left(1-F\left(\frac{m_c^2}{m_b(\nb\cdot q+\omega_1)}\right)\right) 
-\nb\cdot q \left(1-F\left(\frac{m_c^2}{m_b\nb\cdot q}\right)\right)\right.\nn\\
&\left.-\nb\cdot q \left( G\left(\frac{m_c^2}{m_b(\nb\cdot q+\omega_1)}\right) - G\left(\frac{m_c^2}{m_b\nb\cdot q}\right)\right)\right]\,.
\end{align}
$C_{\rm OPE}$ is defined via the OPE result of the decay rate $\Gamma_\mathrm{OPE}$.\footnote{The OPE result of the decay rate is given by (see for more details Ref.~\cite{Benzke:2017woq})
\begin{align}
\Gamma_\mathrm{OPE} =&\, \frac{G_F^2\alpha m_b^5}{32\pi^4}\,|V_{tb}^*V_{ts}|^2\frac{1}{3}\frac{\alpha}{\pi}\int\frac{d\bar n\cdot q}{\bar n \cdot q}
\left(1-\frac{\bar n\cdot q}{m_b}\right)^2\nonumber\\
&\,\Bigg[ C_{7\gamma}^2\Bigg(1+\frac{1}{2}\frac{\bar n\cdot q}{m_b}\Bigg)
+(C_9^2+C_{10}^2)\Bigg(\frac{1}{8}\frac{\bar n\cdot q}{m_b}+\frac{1}{4}\left(\frac{\bar n\cdot q}{m_b}\right)^2\Bigg)
+C_{7\gamma}C_9\frac{3}{2}\frac{\bar n\cdot q}{m_b}\Bigg]\nonumber\\
\equiv&\,\frac{G_F^2\alpha m_b^5}{32\pi^4}\,|V_{tb}^*V_{ts}|^2\frac{1}{3}\frac{\alpha}{\pi}\, C_{\rm OPE}\,.\nonumber
\end{align}
}
$F(x)$ is the penguin function defined in the previous section. The second penguin function is given by $G(x)=2\sqrt{4x-1}\arctan(1 / \sqrt{4x-1}) - 2$. 

For the analysis of the resolved contribution from the interference of ${\cal O}_1$ and ${\cal O}_7$ in the case of $\bar B \to X_{s} \ell^+\ell^-$ we follow the same strategy as in the case of $\bar B \to X_{s} \gamma$ and use the same basis of functions. {We also take the Wilson coefficients in the resolved contributions at the hard scale as our default value and explore the scale dependence by running down to the hard-collinear scale.  The hard scale is the natural choice for the OPE results.} 
We also use the same grid of input parameters and make a multi-parameter scan to find the extreme values 
of the convolution integral.

There are two features which  are crucial to understand our results which we present below. 
\begin{itemize}
\item First, due to the rather symmetric structure of the jet functions, in contrast to the $\bar B \to X_s \gamma$ case, the various model functions lead to very similar extreme values of the convolution integral as we will see below. This feature is already manifest in the bottom of Figure~\ref{Fig:bsll}, where some model functions are shown. Thus, using higher-order polynomials does not increase the uncertainties compared to the second-order polynomial used in the original analyses. 
\item Second, in the upper plot of Figure~\ref{Fig:bsll}, two input values of the jet function, namely the charm and the bottom masses, $m_c$ and $m_b$, are varied within their $1 \sigma$ uncertainties. As in the case of  $\bar B \to X_{s} \gamma$   one finds that larger $m_c$ and smaller $m_b$ values move the jet function to the right, outside the hadronic range. 
Thus, as in the case of $\bar B \to X_s \gamma$ the convolution with the shape functions leads to larger values, if $m_c = 1.14$\, and $m_b =4.61$\,GeV.  
{However, in contrast to the $\bar B \to X_s \gamma$ case, the jet function 
{has a comparatively broad peak.}
Therefore the variation of the charm mass has a lower impact on the magnitude of the 
convolution integral in the $\bar B \to X_{s} \ell^+\ell^-$ case. }
\end{itemize}
\begin{figure}[hpt]
 \includegraphics[scale=1.00]{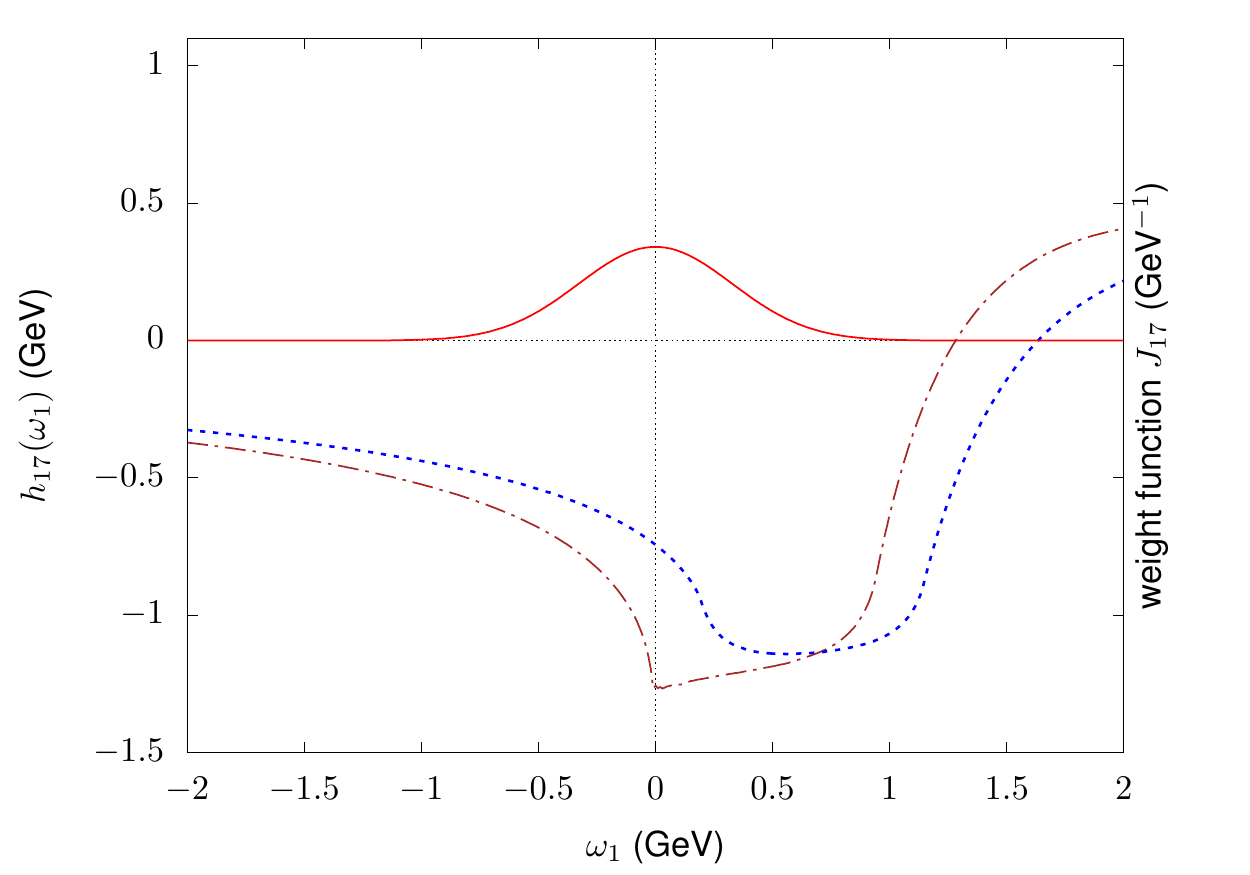} \includegraphics[scale=1.00]{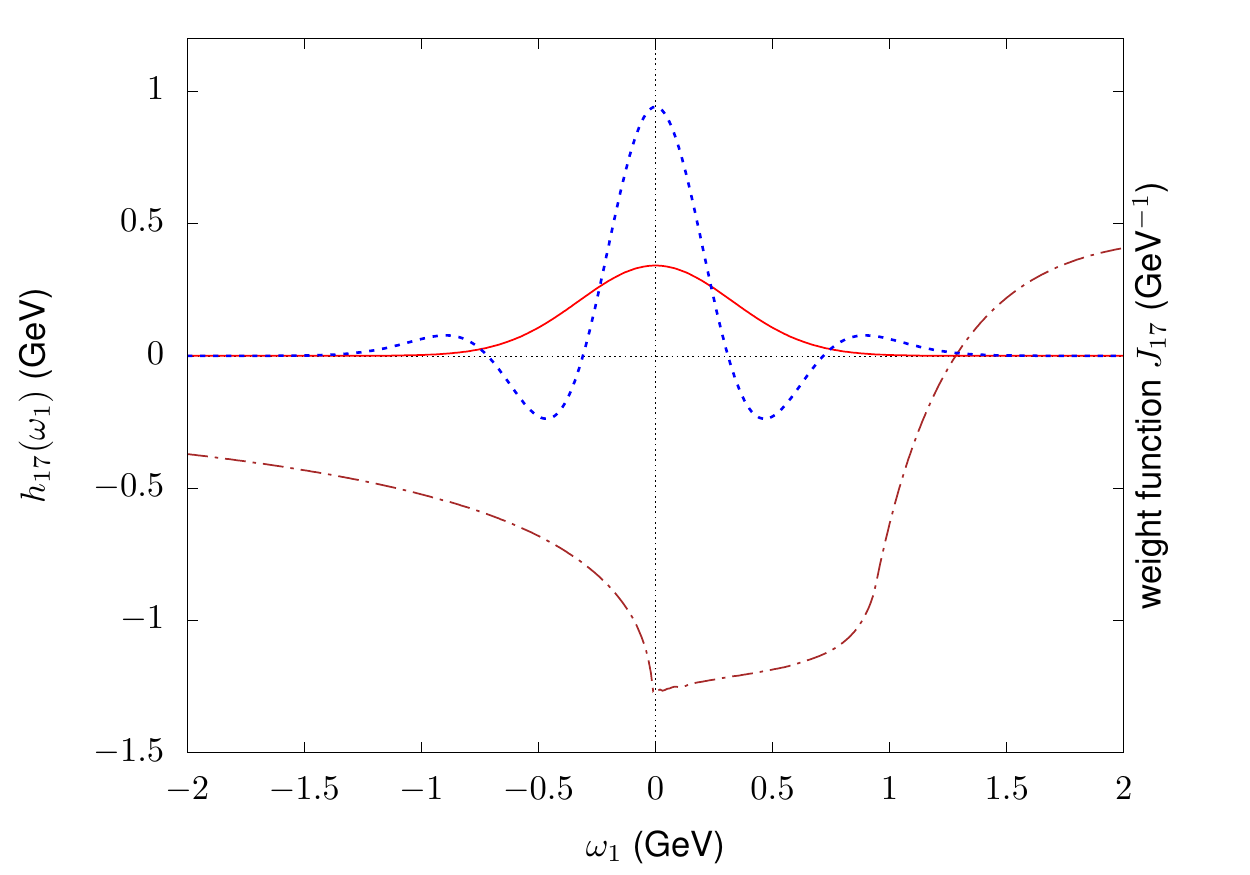}
 \caption{The top figure shows the jet (weight) function in the case $\bar B \to X_s \ell^+\ell^-$ for $m_c= 1.14$\,GeV and $m_b = 4.61$\,GeV (dashed-dotted, brown) 
and for $m_c= 1.23$\,GeV and $m_b = 4.55$\,GeV (dotted, blue) with a second order polynomial as shape function (solid, red). The bottom figure shows two shape functions which lead to the extreme values for the convolution. The polynomials are of order two (solid, red) and 
of order 4 (dotted, blue). \label{Fig:bsll}}
\end{figure}
{In order to systematically compare our results we define the parameter $\Sigma_{17}$  in view of Eq.~(\ref{eqn:bsll})) via 
\begin{equation}
{\mathcal F}^{17}_{\rm b \to s \ell \ell} = \frac{1}{m_b}\frac{C_1(\mu)C_{7\gamma}(\mu)}{C_{\rm OPE}} \, \Sigma_{17}\,,  \label{relative uncertaintybsll}
\end{equation}
analogously to Eq.~(\ref{relative uncertainty}).
Starting with the sum of Hermite polynomials of $n=0$ and $n=2$ (see Eq.~\ref{Hermite}) as model function for $h_{17}$ we find in our multi-parameter scan 
\begin{equation}
-195\,\text{MeV} \leq \Sigma_{17} \leq - 48\,\text{MeV}\hspace{1cm}(n \leq 2,{\rm exp}(-x^2)).
\end{equation} 
The lower bound is found with $\sigma =320\,\text{MeV}$, with the zeroth moment 
$m_0 = 0.200\,\text{GeV}^2$ and with the second moment $m_2 = 0.030\,\text{GeV}^4$. This implies for the higher moments \mbox{$m_4 = 0.009\,\text{GeV} ^6$}
and $m_6 = 0.005\,\text{GeV}^8$. The upper bound corresponds to the parameter set, $\sigma = 360\,\text{MeV}$, $m_0 = 0.200\,\text{GeV}^2$, and 
$m_2 = 0.270\,\text{GeV}^4$. The sum of Hermite polynomials up to order $n=4$ leads to 
 \begin{equation}
- 209\,\text{MeV} \leq \Sigma_{17} \leq - 46\,\text{MeV}\hspace{1cm}(n \leq 4,{\rm exp}(-x^2)).
\end{equation} 
The lower bound corresponds to the parameter set, $\sigma = 300\, \text{MeV}$, $m_0 = 0.280\,\text{GeV}^2$, $m_2 = 0.030\,\text{GeV}^4$, 
 and {$m_4 = 0.040\,\text{GeV}^6$}, the upper bound to $\sigma = 320\, \text{MeV}$, $m_0 = 0.200\,\text{GeV}^2$, $m_2 = 0.270\,\text{GeV}^4$
and $m_4 = 0.180\,\text{GeV}^6$. The sum of Hermite polynomials up to order 6 leads to a slightly larger interval for $\Sigma_{17}$:
 \begin{equation}
- 209\,\text{MeV} \leq \Sigma_{17} \leq - 42\,\text{MeV}\hspace{1cm}(n \leq 6,{\rm exp}(-x^2)).   
\end{equation} 
with the lower bound corresponding to the parameters $\sigma = 280\, \text{MeV}$, $m_0 = 0.280\,\text{GeV}^2$, $m_2 = 0.030\,\text{GeV}^4$,
$m_4 = -0.060\,\text{GeV}^6$, and $m_6 = -0.120\,\text{GeV}^8$ and the upper bound to $\sigma = 320\, \text{MeV}$, $m_0 = 0.200\,\text{GeV}^2$, $m_2 = 0.270\,\text{GeV}^4$, $m_4 = 0.240\,\text{GeV}^6$, and $m_6 = 0.280\,\text{GeV}^8$. With an additional polynomial of degree 8 one finds a slightly smaller interval:
 \begin{equation}
- 201\,\text{MeV} \leq \Sigma_{17} \leq - 43\,\text{MeV}\hspace{1cm}(n \leq 8,{\rm exp}(-x^2)).
\end{equation} 
The lower bound is obtained for $\sigma = 380\, \text{MeV}$, $m_0 = 0.280\,\text{GeV}^2$, $m_2 = 0.030\,\text{GeV}^4$,
$m_4 = 0.060\,\text{GeV}^6$, $m_6 = 0.100\,\text{GeV}^8$, and $m_8 = 0.200\,\text{GeV}^{10}$, the upper bound for $\sigma = 320\, \text{MeV}$, $m_0 = 0.200\,\text{GeV}^2$, $m_2 = 0.270\,\text{GeV}^4$, $m_4 = 0.220\,\text{GeV}^6$, $m_6 = 0.260\,\text{GeV}^8$, and $m_8 = 0.400\,\text{GeV}^{10}$.

As in the case of $\bar B \to X_s \gamma$, we also use model functions with ${\rm exp}(-x^4)$ and ${\rm exp}(-x^6)$ suppression because also those model functions can be expressed in terms of basis of Hermite polynomials with a Gaussian function. In that case we find only slightly larger intervals for $\Sigma_{17}$. 
 \begin{equation}
- 211\,\text{MeV} \leq \Lambda_{17} \leq - 48\,\text{MeV}\hspace{1cm}(n \leq 4,{\rm exp}(-x^4)).
\end{equation} 
The lower bound corresponds to the parameter set, $\sigma = 660\, \text{MeV}$, $m_0 = 0.280\,\text{GeV}^2$, $m_2 = 0.030\,\text{GeV}^4$, $m_4 = 0.040\,\text{GeV}^6$, the upper bound to $\sigma = 800\, \text{MeV}$, $m_0 = 0.200\,\text{GeV}^2$, $m_2 = 0.270\,\text{GeV}^4$ and 
$m_4 = 0.140\,\text{GeV}^6$. 
With the Hermite polynomials up to degree 6 with an ${\rm exp}(-x^4)$ suppression, one obtains the largest interval:
 \begin{equation}
- 215\,\text{MeV} \leq \Sigma_{17} \leq - 36\,\text{MeV}\hspace{1cm}(n \leq 6,{\rm exp}(-x^4)). \label{Maximumbsll}
\end{equation} 
The corresponding parameter sets are $\sigma = 620\, \text{MeV}$, $m_0 = 0.280\,\text{GeV}^2$, $m_2 = 0.030\,\text{GeV}^4$, $m_4 = 0.060\,\text{GeV}^6$, and $m_6 = 0.060\,\text{GeV}^8$ for the lower bound and $\sigma = 760\, \text{MeV}$, $m_0 = 0.200\,\text{GeV}^2$, $m_2 = 0.270\,\text{GeV}^4$, $m_4 = 0.240\,\text{GeV}^6$, and $m_6 = 0.260\,\text{GeV}^8$ for the upper bound.
If one uses a higher suppression, namely ${\rm exp}(-x^6)$ for example with a Hermite polynomial up to degree 4, one already gets a slightly smaller interval again, namely 
 \begin{equation}
- 215\,\text{MeV} \leq \Sigma_{17} \leq - 52\,\text{MeV}\hspace{1cm}(n \leq 4,{\rm exp}(-x^6))\,,
\end{equation} 
with $\sigma = 720\, \text{MeV}$, $m_0 = 0.280\,\text{GeV}^2$, $m_2 = 0.030\,\text{GeV}^4$, $m_4 = -0.300\,\text{GeV}^6$ for the lower bound and $\sigma = 740\, \text{MeV}$, $m_0 = 0.200\,\text{GeV}^2$, and $m_2 = 0.270\,\text{GeV}^4$. 
$m_4 = 0.200\,\text{GeV}^6$ for the upper bound.

Therefore the largest interval for $\Sigma_{17}$ is again found for a sum of Hermite polynomials up to degree 6 with an ${\rm exp}(-x^4)$ suppression, which leads to a range 
$- 215\,\text{MeV} \leq \Sigma_{17} \leq - 36\,\text{MeV}$. However, all the other model functions used above lead to very similar results. Thus, adding higher-order polynomials and using higher suppression factors have almost no effect in the $\bar B \to X_s \ell^+\ell^-$ case in contrast to the $\bar B \to X_s \gamma$ case. This effect can be regarded as a consequence of the rather symmetric jet function as anticipated at the beginning of this section. 
The interval found in the original analysis of $\bar B \to X_s \ell^+\ell^-$ in Ref.~\cite{Benzke:2017woq} was $- 355\,\text{MeV} \leq \Sigma_{17} \leq +50\,\text{MeV}$.\footnote{We note that the factor $e_c$ was not included in $\Sigma_{17}$ in Ref.~\cite{Benzke:2017woq}, so in section 6.1 of that reference one finds the interval $- 532\,\text{MeV} \leq \Sigma_{17} \leq +75\,\text{MeV}$.}
Therefore the size of the interval found in our new analysis is by more than a factor of two smaller. 
\vspace{0.2cm}

Furthermore, as in the case of $\bar B \to X_s \gamma$
there exists an additional $1/m_b^2$ correction in our formula which was neglected in Eq.~\ref{eqn:bsll} at order $1/m_b$. In order to take it into account we have to replace Eq.~\ref{eqn:bsll} by the following original one\footnote{For the precise limits of integration we refer the reader to the discussion in Section 6.1 of Ref.~\cite{Benzke:2017woq}.}
\begin{align} 
&{\mathcal F}_{17}
=\frac{1}{m_b} {\frac{C_1(\mu)C_{7\gamma}(\mu)}{C_{\rm OPE}}}\, e_c\,
\mathrm{Re} \int_{-\infty}^{+\infty}\frac{d\omega_1}{\omega_1+\ie}\, \int \frac{d\nb\cdot q}{\nb\cdot q}\, \int d\omega\,  \frac{(m_b + \omega)^3}{m_b^3}\nn\\
&\frac{1}{\omega_1}\left[
(\nb\cdot q+\omega_1)\left(1-F\left(\frac{m_c^2}{(m_b + \omega)\,(\nb\cdot q+\omega_1)}\right)\right) 
-\nb\cdot q \left(1-F\left(\frac{m_c^2}{(m_b + \omega)\, \nb\cdot q}\right)\right)\right.\nn\\
&\left.-\nb\cdot q \left( G\left(\frac{m_c^2}{(m_b + \omega)\,(\nb\cdot q+\omega_1)}\right) - G\left(\frac{m_c^2}{(m_b + \omega)\,\nb\cdot q}\right)\right)\right]
g_{17}(\omega,\omega_1,\mu)\,.
\label{eqn:f17}
\end{align}
If we include the $1/m_b^2$ term we again find the extrema for $\Sigma_{17}$ for almost the same parameters as in the corresponding cases 
without the $1/m_b^2$ correction. Using a Gaussian suppression in the model function the largest interval is found for the sum of Hermitian polynomials up to degree 6 which leads to the largest interval: 
\begin{equation}
- 259\,\text{MeV} \leq \Sigma_{17} \leq -30\,\text{MeV}\,.
\end{equation}

If one chooses an ${\rm exp}(x^{-4})$ suppression, the polynomial of degree 6 leads to the maximal result
 \begin{equation}
- 268\,\text{MeV} \leq \Sigma_{17} \leq -18\,\text{MeV}\,. \label{Maximum2bsll}
\end{equation} 
We note that this $1/m_b^2$ effect which belongs to the ${\cal O}_1 - {\cal O}_{7\gamma}$ contribution was not included in the original analysis in Ref.~\cite{Benzke:2017woq}.

{{Finally, the shape functions which lead to extreme convolutions with the jet functions do all have relatively small higher moments because
large higher moments correspond to shape functions with maxima close to the hadronic limits.
Therefore the dimensional estimates on the fourth and sixth  moments, given in Eq.~(\ref{highermoments}), namely that their  values  are between $-0.3\,{\rm GeV^6}$ and $0.3\,{\rm GeV^6}$ and between $-0.3\,{\rm GeV^8}$ and $0.3\,{\rm GeV^8}$, respectively, have almost no impact on the results in the case of the decay $\bar B \to X_s \ell^+\ell^-$ because these constraints are automatically fulfilled in almost all cases due to the symmetric jet function.} {Only  the model function with $n \leq 6$ and ${\rm exp}(-x^4)$ which leads to the largest interval would allow for even larger values when the dimensional estimates were not used; the upper bound would slightly move up  from 
$- 18\,\text{MeV}$ to $- 6\,\text{MeV}$ (with the $1/m_b^2$ correction included)}.  {In contrast, the jet function in the $\bar B \to X_s \gamma$ case is peaked and
asymmetric; thus, maxima of the shape function at the border of the hadronic range lead to larger convolutions with this jet function and this leads to larger higher moments of the shape functions. This explains the large impact of the additional estimates of  the fourth and sixth moment found in the $\bar B \to X_s \gamma$ case.}}\\

{\bf Summary of numerical results in the case of $\bar B \to X_{s,d}  \ell^+\ell^-$:}
We found the new {conservative estimate} for $\Sigma_{17}$ at order $1/m_b$ given  in  Eq.~\ref{Maximumbsll}\,, namely $- 220\,\text{MeV} \leq \Sigma_{17} \leq - 40\,\text{MeV}$. This result translates into the following relative uncertainty of the decay rate of $\bar B \rightarrow X_s \ell^+\ell^-$ via Eq.~\ref{relative uncertaintybsll}:
\begin{equation}
{\cal F}_{\rm b \to s \ell\ell}^{17} |_{1/m_b} \in [+0.4 \%,\,+2.1\%]\,,  
\end{equation}
which is more than a factor of two smaller than the uncertainty of our original analysis in Ref.~\cite{Benzke:2017woq}, namely ${\cal F}_{\rm b \to s \ell\ell}^{17} |_{1/m_b} \in [-0.5\%,\,+3.4\%]  $. Including the large additional $1/m_b^2$ contribution, given in Eq.~\ref{Maximum2bsll}\,,\, 
$- 270\,\text{MeV} \leq \Sigma_{17} \leq - 20\,\text{MeV}$,
we arrive at our final result:
\begin{equation}
 {\cal F}_{\rm b \to s \ell\ell}^{17} \in [+0.2\%,\,+2.6\%] \,.   
\end{equation}
 
Our results are rather independent from the specific choice of the degree of the polynomial and of the suppression function used. 
Moreover, the dimensional estimates on the fourth and sixth moments in Eqs. (\ref{highermoments}) have almost no impact on our result in the $b \to s \ell\ell$ case in contrast to the $b \to s \gamma$ case {We showed that both features are consequences of the specific form of the jet functions.}

{Regarding scale variations in our final result, all remarks made in the  $\bar B \to X_s\gamma$ case also apply in this case.} 

The two other resolved contributions at order $1/m_b$ due to the interference ${\cal O}_{7\gamma} - {\cal O}_{8g}$ and ${\cal O}_{8g} - {\cal O}_{8g}$ were estimated in our original analysis in ref.~\cite{Benzke:2017woq} to 
$ {\cal F}_{\rm b \to s \ell\ell}^{78} = [0\%,\,0.1\%] $  and $ {\cal F}_{\rm b \to s \ell\ell}^{88} = [0\%,\,0.5\%]$, respectively. 
Adding the three contributions by using the scanning method, we arrive at the final result for all resolved contributions at order $1/m_b$ (including the additional $1/m_b^2$ piece within ${\cal F}^{17} ):$
\begin{equation} 
{\cal F}_{\rm b \to s \ell\ell}^{1/m_b} \in  [0.2\%,\,3.2\%]\,.
\end{equation}

As was already emphasised in our original analysis, there are subleading contributions due to the interference of
${\cal O}_{9,10}$ and ${\cal O}_1$ at order $1/m_b^2$ which are numerically relevant due to the large ratio $C_{7\gamma}/C_{9,10}$ and which will be presented in Ref.~\cite{Benzkenearfuture}. 

The necessary modifications for the $\bar B \to X_d \ell^+\ell^-$ decay can be found in Refs.~\cite{Huber:2019iqf,Hurth:2018ryv}.

\section{Final summary and conclusions} \label{sec:conclusion}

The nonlocal power corrections to the decays $\bar B \to X_s \gamma$ and $\bar B \to X_{s,d} \ell^+\ell^-$ represent the largest 
uncertainties (around $\pm 5\%$) of the theoretically clean inclusive penguin modes~\cite{Misiak:2015xwa,Huber:2015sra,Huber:2019iqf}. These resolved contributions had been estimated using soft-collinear effective theory (SCET) for the $\bar B \to X_s \gamma$ in Ref~.\cite{Benzke:2010js} and for the $\bar B \to X_{s} \ell\ell$ case in Ref.~\cite{Benzke:2017woq}.
The largest resolved contribution in both cases is due to the interference of the effective operators ${\cal O}_1$ and ${\cal O}_{7\gamma}$. 

The resolved contributions are given by convolution integrals of a so-called jet function, characterising the hadronic final state $X_{s}$ at the intermediate hard-collinear scale $\sqrt{m_b \Lambda_{\rm QCD}}$, and of a soft (shape) function at scale $\Lambda_{\rm QCD}$ which is defined by an explicit non-local heavy-quark effective theory (HQET) matrix element while the hard contribution at the scale $m_b$ is factorised into the Wilson coefficients.
Knowing the explicit form of the HQET matrix element one derives general properties of this shape function and uses model functions with all these properties to estimate the convolution integral with the perturbatively calculable jet function.

In the two original analyses of the most important resolved contribution of 
${\cal O}_1 - {\cal O}_{7\gamma}$~\cite{Benzke:2010js,Benzke:2017woq} only polynomials of second order with a Gaussian suppression were used as model functions for the shape functions. Their parameters were scanned in order to find the most conservative estimate for the convolution integral with the corresponding jet functions.

In a recent analysis in Ref.~\cite{Gunawardana:2019gep} the authors offered a reevaluation of this resolved contribution in the case of $\bar B \to X_s \gamma$. 
They derived a new constraint on the second moment of the corresponding shape function and then made a systematic analysis of model functions based on a complete basis of functions using the Hermite polynomials as was already advocated and used in several applications by the authors of Refs~\cite{Ligeti:2008ac,Lee:2008xc,Bernlochner:2020jlt}. 
{This systematic approach allows to avoid any prejudice  regarding the unknown functional form of the shape functions.}. Using additional dimensional estimates on the fourth and sixth  moment, the authors of Ref.~\cite{Gunawardana:2019gep} found the uncertainty due to this
resolved contribution of ${\cal O}_1 - {\cal O}_{7\gamma}$ reduced by a factor of three. 

In our present analysis of this resolved contribution to the $\bar B \to X_s \gamma$ and also to the $\bar B \to X_s \ell^+\ell^-$ decay, we followed the same strategy of a systematic analysis and also used {the constraint on the second 
moment.}  In addition we analysed the impact of the dimensional estimates of the fourth and  the sixth moment derived in Ref.~\cite{Gunawardana:2019gep}. We found a  significantly smaller  reduction in the case $\bar B \to X_s \gamma$ and a reduction by a factor of two in the case $\bar B \to X_s \ell^+\ell^-$. We explicitly worked out the differences of our result compared to the one of recent analysis of the $\bar B \to X_s \gamma$ case in Ref.~\cite{Gunawardana:2019gep}:
First, we included  the very large $1/m_b^2$ contribution which directly originates  from the resolved contribution ${\cal O}_1 - {\cal O}_{7\gamma}$ and which was also included in the original analysis in Ref.~\cite{Benzke:2010js}. 
Other resolved $1/m_b^2$ contributions like the ones  due to  the operator pairs ${\cal O}_1 -  {\cal O}_{8 g}$  or ${\cal O}_1 - {\cal O}_{1}$  were shown to be numerically negligible in the original analysis. However, the $1/m_b^2$  term in   ${\cal O}_1 - {\cal O}_{7\gamma}$  was dropped in the recent analysis in Ref.~\cite{Gunawardana:2019gep}.  Second, we take into account a larger uncertainty due to the charm mass.  {These two differences have the largest impact.} 
Third, we explore the full space of functions. given by the Hermite polynomials and also used polynomials with suppression factors ${\rm exp}(-x^4) $ or ${\rm exp}(-x^6)$. Such functions can be expressed in terms of the  original basis given in Eq.~\ref{Hermite}.
Fourth, we use a more dense parameter grid in our analysis. {If one does not assume  the dimensional estimates 
on the fourth and sixth moment, we find significantly larger values for the resolved contributions  which shows the large  
impact of these dimensional estimates.}

In contrast to the $\bar B \to X_s \gamma$ case we found that the additional constraint on the second moment -- established in the recent analysis in Ref.~\cite{Gunawardana:2019gep} -- has a much larger  impact in the $\bar B \to X_s \ell^+\ell^-$ decay. It leads to a reduction of the uncertainty due to ${\cal O}_1 - {\cal O}_{7\gamma}$ by a factor of two compared to the result in our original analysis~\cite{Benzke:2017woq}. We also identified the main reason which lead to these different results in the two penguin modes. The jet function in the $\bar B \to X_s \ell^+\ell^-$ case is symmetric and has a broad peak, while the jet function in the   $\bar B \to X_s \gamma$ case is asymmetric and peaked. 
 Therefore, the choice of higher-order polynomials has no impact on the convolution integral in contrast to the  $\bar B \to X_s \gamma$ case.  The special features of the jet function in the $B \to X_s \ell^+\ell^-$ case also implies 
that the charm dependence is less pronounced and that  {the dimensional constraints}  on the fourth and sixth  moments on the shape function have no impact either. Finally, we mention that we also estimated the large $1/m_b^2$ term in the ${\cal O}_1 - {\cal O}_{7\gamma}$ contribution to the $\bar B \to X_s \ell^+ell^-$ decay which we now included in the final result.
 
We found a large scale ambiguity in the final results {which was never explicitly addressed in previous work.} The only scale in our resolved contribution is within the hard function, represented by the Wilson coefficients. Therefore we have chosen the hard scale for the Wilson coefficients as our default value.  If we run down the LO Wilson coefficients in the resolved contribution, i.e.   $C_1(\mu)$, $C_{7\gamma}(\mu)$ in the ${\cal O}_1$-${\cal O}_{7\gamma}$ term, to the hard-collinear scale, the result increases by more than  $40\%$. There is no strict argument here that this specific scale variation in our result  can be connected to an estimate of the unknown NLO corrections. However,  this observation calls for a calculation of the $\alpha_s$ corrections and RG resummation. We found that the charm dependence of our result in the $\bar B \to X_s\gamma$ case is very pronounced. A calculation of the $\alpha_s$ corrections would also allow to control the charm mass dependence of our result.

We conclude that the nonperturbative nonlocal corrections to the $\bar B \to X_s \gamma$ decay still represents the largest uncertainty in this decay mode.
In the case of the $\bar B \to X_s \ell^+\ell^-$ decay we found a reduction of the uncertainty by factor of two due to the new second moment constraint at order $1/m_b$. However, the calculation of the relevant resolved contributions to the 
$\bar B \to X_s \ell^+\ell^-$ is not complete yet. There are subleading contributions 
due to the interference of
${\cal O}_{9,10}$ and ${\cal O}_1$ 
at order $1/m_b^2$ which are numerically relevant due to the large ratio $C_{7\gamma}/C_{9,10}$ and which will be presented in Ref.\cite{Benzkenearfuture}.

As already discussed by the authors of Ref.~\cite{Gunawardana:2019gep},  further improvements might be possible in the near future. More accurate and new determinations of HQET parameters using future data of the Belle-II experiment and lattice QCD will allow to determine the moments of the subleading shape function $h_{\rm 17}$ more accurately and will allow to reduce the error due the resolved contributions within the two inclusive penguin decays. However, this is a difficult task because determinations of higher moments rely on the so-called Lowest-Lying State Approximation (LLSA).

\section*{Acknowledgement}
We thank Jens Erler, Tobias Huber, Thomas Mannel, and Matthias Neubert for valuable help and Maria Vittoria Garzelli, 
Paolo Gambino, Iain Stewart, Sascha Turczyk, and Frank Tackmann for useful discussions. The work was supported by the Cluster of Excellence ``Precision Physics, Fundamental Interactions, and Structure of Matter" (PRISMA$^+$ EXC 2118/1) funded by the German Research Foundation (DFG) within the German Excellence Strategy (Project ID 39083149). TH thanks the  2nd Institute  for Theoretical Physics at Hamburg University  as well as the CERN theory group for their  hospitality during his regular visits to Hamburg and CERN where part of this work was written. MB is grateful to the Mainz Institute for Theoretical Physics (MITP) for its hospitality and its partial support during the completion of this work.

\end{document}